\documentclass[apj]{emulateapj}
\usepackage{refcount}
\usepackage{amsmath}
\usepackage{amssymb}
\usepackage{hyperref}
\usepackage{euscript}
\usepackage{algorithm2e,algorithmic}
\usepackage{bbold}

\usepackage{paralist}

\def\gtorder{\mathrel{\raise.3ex\hbox{$>$}\mkern-14mu
             \lower0.6ex\hbox{$\sim$}}}
\def\ltorder{\mathrel{\raise.3ex\hbox{$<$}\mkern-14mu
             \lower0.6ex\hbox{$\sim$}}}

\slugcomment{Draft of \today}

\shortauthors{Soumagnac \& Ofek}

\begin{document}

\title{catsHTM - A tool for fast accessing and cross-matching large astronomical catalogs}
\author{Maayane T. Soumagnac\altaffilmark{1} \& Eran O. Ofek\altaffilmark{1}}
% * <maayane.soumagnac@weizmann.ac.il> 2018-02-22T14:42:46.716Z:
%
% ^.
\altaffiltext{1}{Benoziyo Center for Astrophysics, Weizmann Institute
  of Science, 76100 Rehovot, Israel, \\Corresponding author:\\ Maayane T. Soumagnac, maayane.soumagnac@weizmann.ac.il}

\begin{abstract}

Fast access to large catalogs is required for some astronomical applications.
Here we introduce the {\tt catsHTM} tool, consisting of several large catalogs reformatted into  HDF5-based file format, which can be downloaded and used locally. To allow fast access, the catalogs are partitioned into hierarchical triangular meshes and stored in HDF5 files. Several tools are provided to perform efficient cone searches at resolutions spanning from a few arc-seconds to degrees, within a few milliseconds time.
The first released version includes the following catalogs (by alphabetical order):
2MASS, 2MASS extended sources, AKARI, APASS, Cosmos, DECaLS/DR5, FIRST, GAIA/DR1, GAIA/DR2,
GALEX/DR6Plus7, HSC/v2, IPHAS/DR2, NED redshifts, NVSS, Pan-STARRS1/DR1, PTF photometric catalog, ROSAT faint source, SDSS sources,
SDSS/DR14 spectroscopy, Spitzer/SAGE, Spitzer/IRAC galactic center, UCAC4, UKIDSS/DR10,
VST/ATLAS/DR3, VST/KiDS/DR3, WISE and XMM.
We provide {\tt Python} code that allows to perform cone searches,
as well as {\tt MATLAB} code for performing cone searches,
catalog cross-matching, general searches, as well as load and create these catalogs.

\end{abstract}

%methods: statistical ---
%techniques: image proccessing ---
%techniques: photometric}

\section{Introduction}
\label{sec:Introduction}

In the past three decades, the emergence of catalog services like
SIMBAD\footnote{\url{http://simbad.u-strasbg.fr/simbad}} \citep{2000A&AS..143....9W}, VizieR\footnote{\url{http://vizier.u-strasbg.fr}} \citep{2000A&AS..143...23O}, the NASA Extragalactic Database (NED)\footnote{\url{https://ned.ipac.caltech.edu}} and MAST\footnote{\url{https://archive.stsci.edu/astro}}, have had, and continue to have, an enormous impact on astronomical research. These services are being used extensively by the astrophysics community and are used in a large fraction of  articles.
Nevertheless, one obvious limitation of all these services is that the Internet connection limits the speed of search. This is an obstacle for some applications requiring very fast access to large catalogs.
Examples of such applications are cross-matching multiple large catalogs
(e.g., \citealt{2012BaltA..21..319M,2015ASPC..495...25O}), and vetting transient candidates detected by synoptic surveys.
For example, surveys like Pan-STARRS \citep{2016arXiv161205560C} and the Palomar Transient Factory (Law et al. 2009)
generated a large number of transient candidates which have to be cross-matched with multiple catalogs in order to be vetted, classified and followed up. 

Cross-matching tools developed in the last years include the web-based tools CDS-Xmatch \citep{2011ASPC..442...85P} and ARCHES \citep{2015ASPC..495..437M}, or local command line tools like TOPCAT \citep{2005ASPC..347...29T}, STILTS \citep{2006ASPC..351..666T} and $C^3$ \citep{2017PASP..129b4005R} which allow to overcome some of the speed limitations of the web-based applications. Another simple solution is to use local relational databases. Indeed, with proper indexing (e.g., Hierarchical Triangular Mesh (HTM); \citealt{2007cs........1164S}),
Structured Query Language (SQL) queries on such database are relatively fast. In this paper we present a simple and - at least from our experience - faster alternative.

We provide a set of large catalogs stored in HDF5 files\footnote{\label{hdf5}\url{https://support.hdfgroup.org/HDF5/}}. HDF5 is a data model, library and file format for storing and managing data. It supports an unlimited variety of data types and is designed for flexible and efficient I/O and for high volume and complex data. Furthermore, HDF5 tools are available in many computer languages. The data storage methodology we use is designed to provide good performances both for small size (i.e., a few arcsec) and large size (i.e., deg) cone searches. In addition to the formatted catalogs, we provide a set of tools to perform fast cone search, serial search, catalog cross-matching and catalog generation.

The structure of this paper is as follows.
In \S\ref{sec:Format}, we detail the structure and format of the catalogs. In \S\ref{sec:Cat} we list all the catalogs currently available and present the codes we provide for fast access to these catalogs. We compare our tool to other existing tools and discuss our results in \S\ref{sec:disc}.

\section{The data format}
\label{sec:Format}

The efficiency of {\tt catsHTM} lies in three aspects: (1) the way the data is partitioned into files;
(2) the way the data is stored in these files;
and (3) the data indexing.

\begin{figure}
\centerline{\includegraphics[width=6cm]{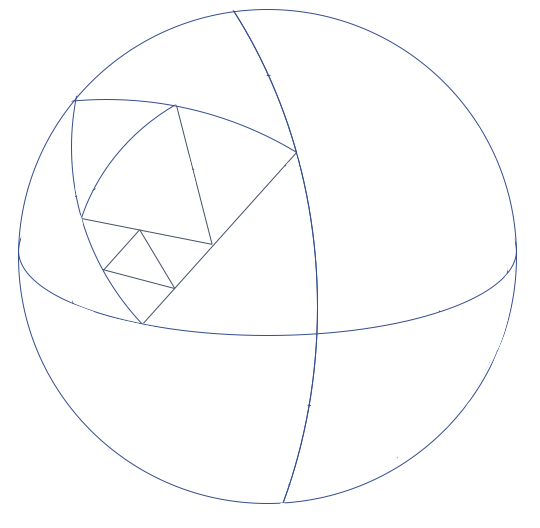}}
\caption{Illustration of the recursive decomposition of the sphere into trixels, up to level $3$. The trixels are spherical triangles: polygons of which the edges are segments of three great circles. Each trixel has four "sons" (i.e.; quad tree).}
\label{fig:htms}
\end{figure}

The data of each catalog is partitioned in the following way. We divide the celestial sphere using a Hierarchical Triangular Mesh (HTM) quad-tree \citep{2007cs........1164S,2000ASPC..216..141K}. This method is based on a recursive subdivision of the celestial sphere into spherical triangles\footnote{A spherical triangle is a polygone of which the edges are segments of three great circles.} of similar shapes, called {\it trixels}. The HTM method of dividing the sphere is particularly good at supporting searches at a wide range of resolutions (hemispheres to arc seconds). In our case, depending on the size of the catalog, the level of the HTM (i.e., the number of subdivisions necessary to create the smallest trixel) varies between six and nine.
%The reason why we do~not use higher levels is that having the data stored continuously in memory improves the efficiency of wide-cone searches, while it has minor impact on narrow-cone queries. 
The amount of levels we chose provides good efficiency for cone searches at resolutions spanning from a few arc-seconds to a degree -- i.e., the typical resolutions usually used for astronomical applications. Each trixel in the quad-tree is allocated a number, which is referred to as the {\it trixel index} throughout this paper. In figure~\ref{fig:htms}, we show an illustration of the recursive decomposition of a sphere into HTM trixels.

To store the data of the partitioned catalogs, we use datasets, one of the object types available within the HDF5 data model\footnotemark[\getrefnumber{hdf5}]. A dataset is simply a multidimensional array and a HDF5 file can contain multiple datasets. 

For each trixel, we create a pair of datasets: 
\begin{enumerate}
\item The {\it trixel-dataset}, containing the actual catalog data that are within the trixel. Only highest level trixels are populated (and exist). The data in each trixel is sorted by declination.
\item The {\it index-dataset}, containing a two column matrix. The first column is the line number in the trixel-dataset and the second column is the corresponding declination. This data exist in steps ranging from $30$ to $300$ lines and can be used in some cases to expedite a search within a trixel. This is mainly used for catalogs cross-matching.
\end{enumerate}
These pairs of datasets are then stored, in groups of 100 (by default), in HDF5 files.

For each partitioned catalog, we create an additional HDF5 file, called the {\it HDF5 index file}, which contains the information needed in order to access the relevant trixels, when making a cone search. The HDF5 index file contains as many lines as trixels in the tree. Each line corresponds to one trixel and contains: the index of the trixel; the indexes of the parents and sons trixels; the sky coordinates (latitude and longitude) of the poles of the great circles that defines the trixel; the number of sources it contains ({\tt NaN} if the trixel does not belong to the highest level in the tree). The HDF5 index file allows to perform an efficient tree search for all the trixels that intersect -- or are contained within -- a given cone.
Finally, we also keep a file containing additional meta data, such as the catalog column names and units.

To summarize: each catalog is partitioned into trixels. For each partitioned catalog, the following files exist: (1) HDF5 files containing datasets where the catalog data can be stored (2) a HDF5 file called the index file, where the index and sky coordinates of each trixel are stored (3) a file containing meta data on the catalog.

In Table~\ref{tab:format}, we present the default files and datasets naming formats we used.
Additional documentation and examples are available on-line\footnote{\label{url}\url{https://webhome.weizmann.ac.il/home/eofek/matlab/doc/catsHTM.html}}.
\begin{deluxetable*}{llll}
\tablecolumns{3}
\tablewidth{0pt}
\tablecaption{Default Naming format}
\tablehead{
\colhead{Object}     &
\colhead{Default naming format}       &
\colhead{type}     
}
\startdata
Trixel-dataset &  htm\_\%06d& HDF5 dataset\\
Index-dataset & htm\_\%06d\_Ind& HDF5 dataset\\
HDF5 files containing the trixel-datasets and the index-datasets &$<$CatBaseName$>$\_htm\_\%06d.hdf5 & HDF5 file\\
HDF5 index file &  $<$CatBaseName$>$\_htm\_Ind.hdf5&HDF5 file\\
Catalog metadata (e.g., column names)& $<$CatBaseName$>$\_htmColCell.mat& .mat file
%Number of pairs of trixel- and index-datasets in HDF5 files & 100 \\
\enddata
\tablecomments{``$<$CatBaseName$>$'' is the catalog name (see Table~\ref{tab:cata}). The default number of pairs of trixel- and index-datasets in each HDF5 file is 100. In the HDF5 index file, the catalog file name index is rounded such that the last two significant digits are always 0 (in case of 100 trixels per HDF5 file).
\label{tab:format}}
\end{deluxetable*}

\section{Available catalogs and codes}\label{sec:Cat}

The catalogs currently available are listed in Table~\ref{tab:cata} and the list will be updated regularly in the on-line documentation\footnotemark[\getrefnumber{url}]. The {\tt catsHTM} directory containing the HTM/HDF5 catalogs requires $\sim2.1$\,TB of disk space. Download instructions as well as links to the code used to create those catalogs can be found in the on-line documentation\footnotemark[\getrefnumber{url}].
\begin{figure}
\centerline{\includegraphics[width=8cm]{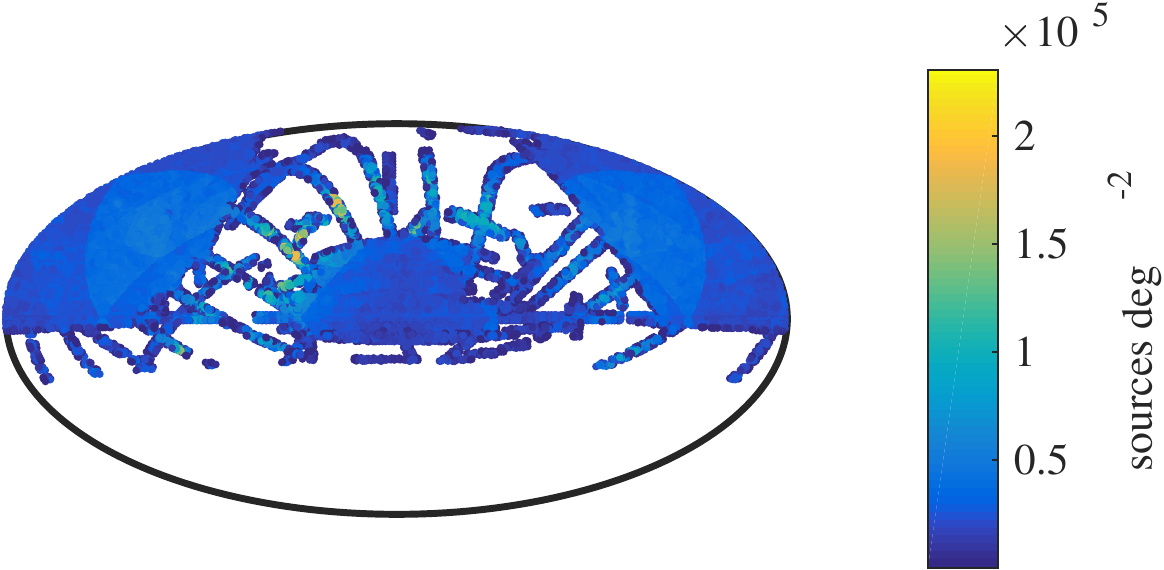}}
\caption{Source density in the SDSS-DR10 catalog as plotted using the \texttt{catsHTM.plot\_density} function.}
\label{fig:AitoffSDSS}
\end{figure}

%APASS\footnote{\label{bla}A newer version of the APASS and NEDz catalogs will be included in one of the next releases.}
%NEDz\footnotemark[\getrefnumber{bla}]

%
\begin{deluxetable*}{lllll}
\tablecolumns{5}
\tablewidth{0pt}
\tablecaption{List of available catalogs (updated in the on-line documentation)}
\tablehead{
\colhead{CatBaseName}     &
\colhead{Catalog}         &
\colhead{$N_{\rm src}$}   &
\colhead{$N_{\rm col}$}   &
\colhead{Reference}      
}
\startdata
TMASS       & 2MASS  &  $470\,992\,608$ & $10$&  \cite{2006AJ....131.1163S}\\
TMASSxsc    & 2MASS extended sources & $1\,647\,599$ &  $15$    &  \cite{2000AJ....119.2498J}    \\
AKARI & AKARI &  $870\,973$     & $9$      & \cite{AKARI}\\
APASS\footnote{\label{bli}A newer version of the APASS catalog will be included in one of the next releases.} & APASS&$55\,395\,532$&$19$&\cite{2009AAS...21440702H}     \\
Cosmos & Cosmos&$438\,226$&$27$&  \cite{2007ApJS..172...99C}     \\
DECaLS & DECaLS/DR5&  $679\,250\,688$    & $26$   & http://legacysurvey.org/decamls/       \\
FIRST & FIRST&  $946\,432$      &  $14$        &   \cite{1995ApJ...450..559B}    \\
GAIADR1 & GAIA/DR1&  $114\,268\,060$     & $8$      &    \cite{GAIA}   \\
GAIADR2 & GAIA/DR2& $1\,692\,967\,552$  & $27$   & \cite{2018arXiv180409365G}   \\
GALEX & GALEX/DR6Plus7&  $165\,794\,048$     & $9$      & \cite{2005ApJ...619L...1M}      \\
HSCv2 & Hubble Source Catalog HSC/v2 &$318\,758\,784$ & $15$ & \cite{2016AJ....151..134W}\\ 
IPHAS & IPHAS/DR2  &   $205\,773\,248$ & $17$ & \cite{2014MNRAS.444.3230B}\\
IRACgc & Spitzer/IRACgc& $1\,065\,565$& $15$ &    \cite{2008ApJS..175..147R}   \\
NEDz & NED redshifts (2018 May 2 version) &        $7\,154\,168$ & $8$ &  \cite{1990ASSL..160..109H,2017IAUS..325..379M}                    \\
NVSS & NVSS& $1\,773\,486$      & $12$      &    \cite{1998AJ....115.1693C}   \\
PS1\footnote{Sources detected in the stacked images above a certain signal-to-noise ratio; the current version of the HDF5 catalog has some missing regions below $\delta=0$.} & Pan-STARRS1/DR1& $2\,612\,435\,712$      &  $41$     &    \cite{2016arXiv161205560C}   \\
PTFpc &PTF photometric catalog &  $21\,167\,678$     &  $14$     &   \cite{2012PASP..124..854O}  \\
ROSATfsc & ROSAT faint source&   $105\,924$    &   $21$    &   \cite{2000yCat.9029....0V}    \\
SAGE & Spitzer/SAGE& $9\,094\,829$& $16$ &  \cite{2006AJ....132.2268M}     \\
SDSSDR10 & SDSS sources&   $447\,279\,840$    & $16$      &  \cite{2014ApJS..211...17A}     \\
SpecSDSS & SDSS/DR14 spectroscopy& $4\,311\,570$& $32$ &    \cite{2017arXiv170709322A}   \\
%SDSSoffset & SDSS/DR14 spectroscopy& $164003184$& $34$ &    \cite{2017arXiv170709322A}   \\
%SWIREz  & SWIRE    &                &           &   \cite{2008yCat.2290....0R}                           \\
UCAC4 & UCAC4&   $113\,780\,216$    &  $45$     &   \cite{2013AJ....145...44Z}    \\
UKIDSS & UKIDSS/DR10&   $79\,333\,520$    &  $38$     &  \cite{2007MNRAS.379.1599L}     \\
%USNOB1 & USNO-B1&       &       & \cite{2003AJ....125..984M}      \\
%VISTAviking & VISTA/Viking/DR2&  $34360912$     &  $18$     & \cite{2013Msngr.154...32E,2016yCat.2343....0E}      \\
VSTatlas & VST/ATLAS/DR3& $106\,556\,464$      & $14$      &   \cite{2015MNRAS.451.4238S}    \\
VSTkids & VST/KiDS/DR3& $48\,735\,816$      & $21$      &    \cite{KIDSDR3}   \\
WISE & WISE      & $563\,908\,224$      & $37$   & \cite{2010AJ....140.1868W}\\
XMM & XMM&    $727\,790$   &    $14$     & \cite{XMM}      
\enddata
\tablecomments{List of the catalogs available at the date of submission (this list will be regularly updated in the on-line documentation). In some cases, there are small discrepancies in the number of sources (typically $<10^{-5}$) between the version available on VizieR and our HDF5 version. Some of these differences are due to problems in the ingestion process and this will be fixed in future releases.
\label{tab:cata}}
\end{deluxetable*}
%founds the basename in gauss eran/matlab/catsHTM/catsHTM/

%\section{Code}\label{sec:Code}
We provide two sets of codes to access the formatted catalogs listed in Table~\ref{tab:cata}:
\begin{enumerate}
\begin{samepage}
\item A {\tt Python} code\footnote{\url{https://github.com/maayane/catsHTM}} that provide cone search functions;
\item A {\tt MATLAB} code\footnote{\url{https://webhome.weizmann.ac.il/home/eofek/matlab/doc/install.html}} that can perform a variety of tasks in addition to cone search, described in the on-line documentation\footnotemark[\getrefnumber{url}]: general searches, cross-matching, plotting, loading catalogs and generating new catalogs. For example, in Figure~\ref{fig:AitoffSDSS} we show a sky map, generated by one of these functions, that presents the source density in the SDSS catalog.
\end{samepage}
\end{enumerate}

Currently, the Python code is compatible with both {\tt Python 2} (higher than {\tt 2.7.10}) and {\tt Python 3} and requires a small amount of standard basic packages ({\tt numpy}, {\tt scipy}, {\tt math} and {\tt h5py}). The {\tt MATLAB} code is available as part of the MATLAB Astronomy \& Astrophysics
Toolbox\footnote{\url{https://webhome.weizmann.ac.il/home/eofek/matlab}} \citep{2014ascl.soft07005O}, and has been tested with {\tt MATLAB} R2016b. These requirements will be updated, if needed, in the online documentation, as future versions of {\tt catsHTM} are released.

\section{Results and discussion}
\label{sec:disc}

Compared to one specific relational database SQL searches, we find our code to be about one order of magnitude faster for cone searches.
A 10\,arcsec radius cone search, on a random celestial position,
using a Xeon(R) CPU E5-2670 v3 at 2.30GHz machine takes on average about 1-4\,ms using the {\tt MATLAB} tool and 1-40\,ms using the {\tt Python} tool. A 1000\,arcsec radius cone search on the same machine takes on average 4-300\,ms using the {\tt MATLAB} tool and 
2-600\,ms using the {\tt Python} tool, where the exact time depends on the catalog.

Our code can also be used to efficiently cross-match catalogs and perform general searches. These options can use multiple processors.
The cross-matching is performed by loading a single trixel dataset from the first catalog into memory and cross-matching it only with sources that are found in overlapping trixels from the second catalog.
For example, using 24 processors, cross-matching of the 
APASS catalog against itself takes about 160\,s 
while cross-matching the 2MASS catalog against the WISE catalog
takes about 53\,min (without dumping the results). The cross-matching tool is work in progress. In particular, it is currently only available in {\tt MATLAB} and will be available in {\tt Python} in future releases of {\tt catsHTM}. In the future, we plan to extend the format for multiple-epoch catalogs, to add new catalogs and to provide catalogs of cross-matchings between all the catalogs.

Making a fair comparison between the performances of {\tt catsHTM} and those of other available tools is a complex task, because {\tt catsHTM} comes with formatted catalogs.

Web-based tools like OpenSkyQuery \citep{2006ASPC..351..493N} and  CDS-Xmatch \citep{2011ASPC..442...85P}, as well as tools with a strong graphical component such as TOPCAT \citep{2005ASPC..347...29T} consist of portals which have the disadvantage of being relatively hermetic to the user. %In particular (1) they limit the ability of the user to have any control on the code and the computational choices involved and (2) they make it harder for the user to integrate the job into his own automatic pipeline. 
The need to allow access to multiple users comes with its own set of compromises: CDS-Xmatch limits both the disk space and computation time available to the users. 

On the contrary, {\tt catsHTM} has been designed as a stand-alone tool, to avoid the above disadvantages and allow high flexibility. The simple cone search python and {\tt MATLAB} codes can be easily edited according to the user's specific needs and easily integrated in the user's own pipelines. The all-in-one structure of the code and catalogs allows the user to run jobs on his own computer and without relying on internet connection. The $C^3$ tool \citep{2017PASP..129b4005R} has been designed in this spirit, but focuses on cross-matching, whereas {\tt catsHTM} primary task is to perform efficient cone-search queries.

Perhaps the existing tool closest to what {\tt catsHTM} is trying to achieve is STILTS \citep{2006ASPC..351..666T} which, on the one hand, is a powerful stand-alone tool, free of the disadvantages of web-page applications, and on the other hand offers a large variety of catalog-handling functionalities, including cone search. 

{\tt catsHTM} is unique in the sense that it is not limited to the search and cross-matching tools it offers. These tools come together with a large - and constantly updated - set of astronomical catalogs, formatted into the binary HDF5 format.

\acknowledgments

E.O.O. is grateful for the support by
grants from the Israel Science Foundation, Minerva, Israeli ministry of Science, the US-Israel Binational Science Foundation,
and the I-CORE Program of the Planning and Budgeting Committee and The Israel Science Foundation.

M.T.S. acknowledges support by a grant from IMOS/ISA, the Ilan Ramon fellowship from the Israel Ministry of Science and Technology and the Benoziyo center for Astrophysics at the Weizmann Institute of Science. M. T. S would also like to thank Steve Schulze, Barak Zakay, Ronen Tamari and Adam Rubin for useful discussions.

%\appendix

%\section{appendix material}
%\label{Ap:}
%\begin{thebibliography}{dummy}
\bibliographystyle{apj}
\bibliography{bibliograph.bib}
%\end{thebibliography}
\end{document}